\documentclass[twocolumn, superscriptaddress]{revtex4-2}

\usepackage{graphicx}
\usepackage{xcolor}
\usepackage{adjustbox}
\usepackage{hyperref, soul}
\usepackage{amsmath}
\usepackage[capitalize]{cleveref}
\usepackage{amsmath}
\usepackage{tikz} 
\usetikzlibrary{calc}

\hypersetup{
    colorlinks,
    linkcolor={blue},
    citecolor={blue},
    urlcolor={blue}
}

\begin{document}

\title{Error correction on an array of superconducting qubits with defective components}

\author{Julien M. Drouet}
\affiliation{School of Physics, The University of Sydney, Sydney, New South Wales 2006, Australia}
\affiliation{School of Computer and Communication Science, Ecole Polytechnique Fédérale de Lausanne, Lausanne 1015, Switzerland}

\author{Xanda C. Kolesnikow}
\affiliation{School of Physics, The University of Sydney, Sydney, New South Wales 2006, Australia}

\author{Campbell K. McLauchlan}
\affiliation{School of Physics, The University of Sydney, Sydney, New South Wales 2006, Australia}

\author{Georgia M. Nixon}
\affiliation{School of Physics, The University of Sydney, Sydney, New South Wales 2006, Australia}

\author{Seok-Hyung~Lee}
\affiliation{Department of Quantum Information Engineering, Sungkyunkwan University, Suwon 16419, Republic of Korea}

\author{Dominic J. Williamson}
\affiliation{School of Physics, The University of Sydney, Sydney, New South Wales 2006, Australia}

\author{Stephen D. Bartlett}
\email{stephen.bartlett@sydney.edu.au}
\affiliation{School of Physics, The University of Sydney, Sydney, New South Wales 2006, Australia}

\author{Benjamin J. Brown}
\affiliation{IBM Quantum, T. J. Watson Research Center, Yorktown Heights, New York 10598, USA}
\affiliation{IBM Denmark, Sundkrogsgade 11, 2100 Copenhagen, Denmark}

\author{Robin Harper}
\affiliation{School of Physics, The University of Sydney, Sydney, New South Wales 2006, Australia}

\begin{abstract}
A solid-state quantum-computing architecture will require the fabrication of arrays of many coupled qubits. It is inevitable that this process will produce qubits and couplers with varying performance, with some components underperforming due to imperfect fabrication. 
Quantum error-correction requires high-performing components and hence these defects must be dealt with, either by adapting the code to exclude the defects, or by informing the decoder to accommodate defects in post-processing.
Here we implement and compare strategies to operate distance-5 surface codes on a quantum processor consisting of a square-lattice array of 120 superconducting qubits.
We demonstrate a dramatic reduction in the probability of a logical error in a memory experiment by excluding underperforming components, compared with both a standard approach of ignoring defects, and a defect-aware decoding approach. We observe up to $2.8\times$ improvement in logical errors per round when excluding defects compared with the standard defect-ignorant approach ($1.62\%$ compared to $4.49\%$). In contrast, defect-aware decoding gives only modest gains. 
Defects are also expected to be particularly harmful for measurement-based logical operations. Using a stability experiment we show that excluding defects resurrects measurement-based logic gate performance, observing a $6.3\%$ per-round suppression of failure rate when excluding defects, compared to zero suppression otherwise.
Furthermore, we show a further substantial decrease in logical errors when using leakage post-selection in combination with our defect exclusion strategies, resulting in a distance-5 code outperforming the best distance-3 in one basis.
Our experiments therefore give a proof-of-principle demonstration of the essential utility of defect exclusion methods in the scale-up of solid-state quantum computing approaches. 
\end{abstract}

\maketitle

\noindent {\it Introduction.---} Quantum error correction (QEC) provides an avenue to execute error-free quantum computations using physical qubits and gates that experience low-level noise. To realise quantum error-correcting codes and the methods of fault-tolerant quantum computation, we need a practical way of fabricating a redundancy of high-quality physical qubits with suitable connectivity to perform the QEC syndrome extraction circuits reliably. A well-established approach is to manufacture a large array of transmon qubits on a superconducting device with couplers that can entangle neighbouring qubits. Such devices have enabled a number of demonstrations of QEC~\cite{Krinner2022, sundaresan2022, Google2022, zhao2022, Gupta2024, Geher:2025aa, Google24, harper2025characterising, Besedin_2026, lee2026scalable, wang2026super}.

As we continue to produce larger qubit arrays, the occurrence of fabrication errors that produce variability in qubit and component performance will become inevitable and, in general, for larger devices, we should expect to have to accommodate some small fraction of underperforming or defective components. If these underperforming components are particularly noisy, they will dramatically impact the logical error probability of a memory unit. Furthermore, defective components can lead to very unreliable measurements of the parity checks used in error correction, which may also introduce logical errors over the course of a logic gate. Variation in the noise model across the different components of a quantum device and circuit can be accommodated at some level by noise-informed decoders~\cite{lee2026scalable}. However, if some components are too noisy, the theoretical literature has proposed that it becomes favourable to implement active strategies where we modify syndrome extraction circuits to exclude the defective components~\cite{Stace2009thresholds, Auger2017, Strikis2021, Siegel2023adaptivesurfacecode, Lin2024codesign, McLauchlan2024accommodating, Wei2025low, Leroux2025snakes, Debroy2025luciinsurfacecode, mishmash2025excising,  wolanski2026automated, wei2026adaptivedeformationcolorcode,higgott2025handlingfabricationdefectshexgrid,anker2025optimizedmeasurementschedulessurface}, effectively quarantining them from the operational device.

Broadly speaking, defect-exclusion strategies seek to measure the parity checks that an ideal, defect-free device was intended to measure. When some of these parity checks cannot be reliably performed due to defective components, the code and syndrome extraction circuit must be modified locally to avoid using these defects, by defining an alternative set of checks that still yields a valid code.
A common approach is to infer the values of larger parity checks, known as super-stabilizers, from collections of lower-weight check measurements that avoid the compromised components. 
Ongoing progress~\cite{Stace2009thresholds, Auger2017, Strikis2021, Siegel2023adaptivesurfacecode, Lin2024codesign, McLauchlan2024accommodating, Wei2025low, Leroux2025snakes, Debroy2025luciinsurfacecode,  wolanski2026automated, wei2026adaptivedeformationcolorcode,higgott2025handlingfabricationdefectshexgrid, mishmash2025excising,anker2025optimizedmeasurementschedulessurface,kim2026luciibmhardwareerror} has continued to improve these defect-exclusion strategies, designing circuits that make better use of the non-defective components in order to minimize the loss of code distance.

In this work we demonstrate the positive utilisation of these strategies in practice.
We adapt the defect-exclusion circuits of Ref.~\cite{Leroux2025snakes} for managing underperforming components on a distance-5 surface code~\cite{Kitaev1997, Dennis2002} realized on the 120-qubit IBM Nighthawk quantum processor \texttt{ibm\_miami}. 
We find the most marked reduction in logical error per round comes from using active defect exclusion strategies with noise-informed decoding. 
In particular, we see the logical error per round for a memory experiment performed in one basis drop from  $4.49\%$ using a standard defect-unaware circuit to $1.62\%$ when we use defect-exclusion strategies. 
We also interrogate strategies to accommodate underperforming components in fault-tolerant logic gates by means of a stability experiment. 
We observe that active defect exclusion enables logical error suppression through increasing the number of syndrome measurement rounds, where this was not possible with just defect-aware decoding. 
Lastly, by post-selecting on samples where it is likely that our system has remained in the computational subspace throughout the experiment, i.e., by removing leakage events through post-selection, we find that active defect exclusion enables a distance-5 code to outperform the best distance-3 code in a specific logical basis. 
This highlights the potential of combining active defect exclusion with a leakage mitigation strategy during syndrome extraction to achieve scalable fault-tolerant quantum computation in these devices.

\begin{figure*}[t]
\centering
\begin{tikzpicture}
    \node[inner sep=0pt] (img)
        {\includegraphics[width=\textwidth]{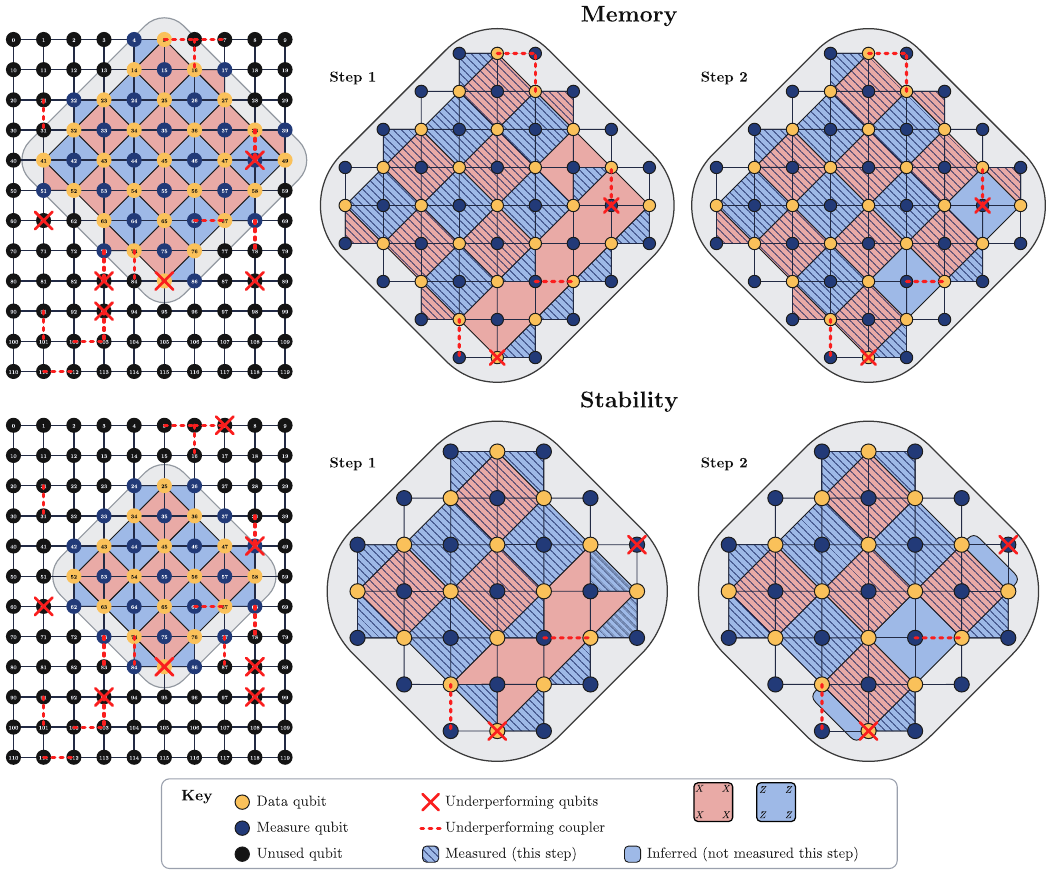}};

    \begin{scope}[
        shift={(img.north west)},
        x={($(img.north east)-(img.north west)$)},
        y={($(img.south west)-(img.north west)$)}]
        \node[anchor=base west] at (-0.005, 0.025) {(a)\hspace{1em}2026-06-10};
        \node[anchor=base west] at (0.310, 0.025) {(b)};
        \node[anchor=base west] at (0.665, 0.025) {(c)};
        \node[anchor=base west] at (-0.005, 0.466) {(d)\hspace{1em}2026-06-17};
        \node[anchor=base west] at (0.310, 0.466) {(e)};
        \node[anchor=base west] at (0.665, 0.466) {(f)};
    \end{scope}
\end{tikzpicture}
    
    \caption{{\bf Surface-code quantum error correction with defect exclusion.} (a)~Arrangement of the 120 qubits (numbered) with couplers (connecting lines between neighbouring qubits) on the IBM Nighthawk quantum processor \texttt{ibm\_miami}.  A distance-5 surface code memory is placed on the device as shown.  Data (measure) qubits are shown in yellow (dark blue). Pauli-X and Pauli-Z stabilizers, illustrated respectively by red and blue plaquettes, can be measured by coupling the measure qubit inside each plaquette to its adjacent data qubits followed by measurement to obtain the parity outcome. Defective components are indicated by red crosses and dashed lines. (b-c)~The two steps of the syndrome extraction process for this defect configuration, excluding defective couplers and measure qubits while retaining the defective data qubit within the code. Dashed plaquettes indicate the stabilizer checks measured during the corresponding step. In step~1, weight-two Pauli-Z operators are measured to infer the value of the weight-four Pauli-Z stabilizers. As these weight-2 Pauli-Z checks do not commute with all of the Pauli-X stabilizers, we form a weight-10 Paul-X super-stabilizer. In step~2 we infer the value of this super-stabilizer using weight-2 and weight-4 Pauli-X checks. (d)~Footprint for a stability experiment. The boundaries are designed such that the product of all Pauli-Z stabilizers, shown in blue, is constrained to return the $+1$ eigenvalue. A sequence of measurement errors on the blue plaquettes over the stability experiment will cause a logical failure (a $-1$ outcome for this constraint). (e-f)~The two-step syndrome extraction measurement sequence for the stability experiment in this defect configuration. The defective couplers and measure qubits are excluded, while the defective data qubit is retained. Dashed plaquettes follow the same convention as in (b-c).}
    \label{fig:layout}
\end{figure*}
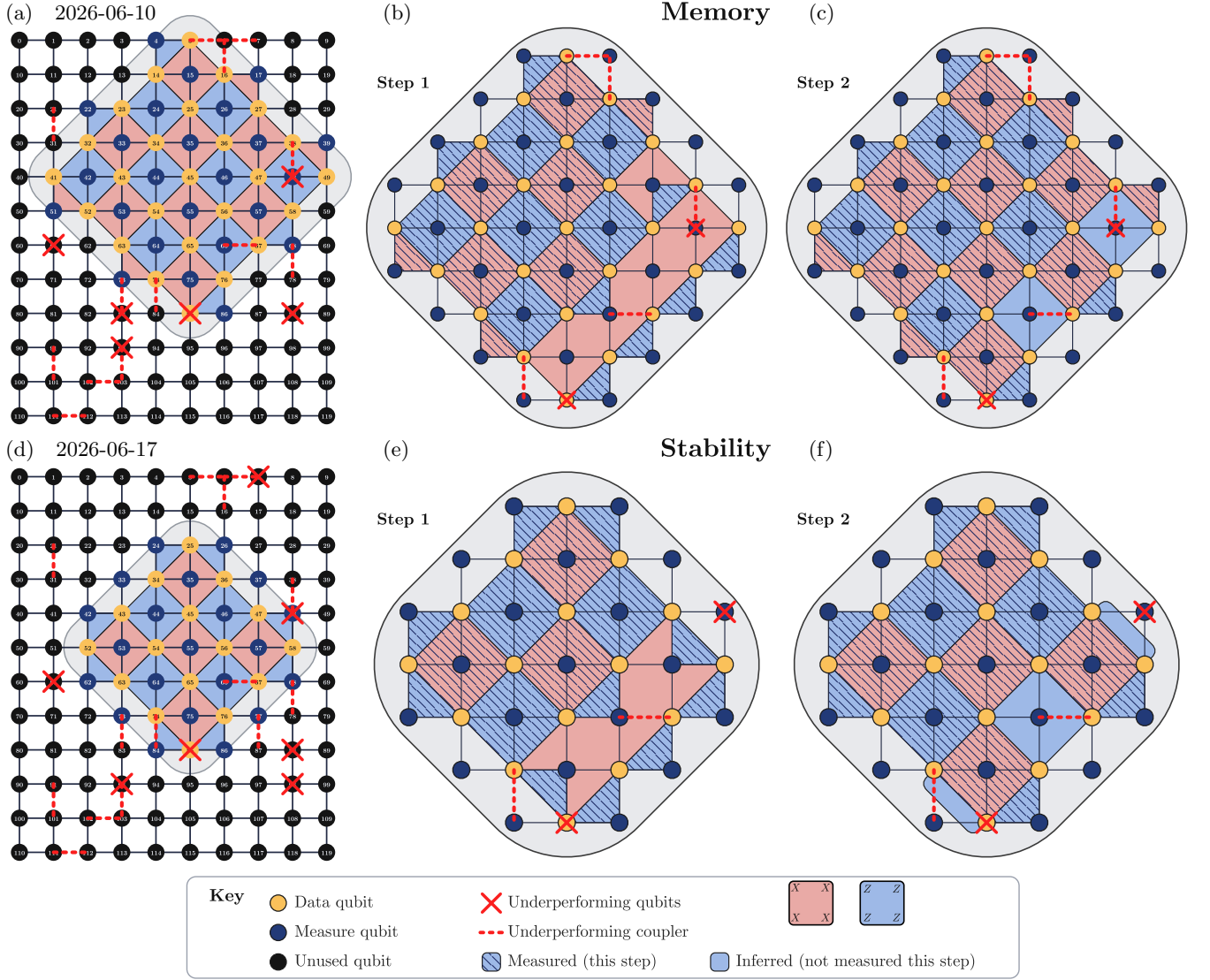

\medskip

\paragraph*{\it Surface-code quantum computing.---}The surface code~\cite{Kitaev1997, Dennis2002, fowler2012b} is most commonly operated on a square array of qubits with nearest-neighbour connectivity. 
Qubits of a square array are partitioned into two subsets: data qubits, and measure qubits. 
Data qubits support the surface code itself, and the additional measure qubits are available to make local parity measurements of stabilizers.  
These stabilizers are commuting observables whose measured values reveal information about the occurrence of errors. 
Typically stabilizers are measured by entangling each measure qubit with its neighbouring data qubits, and then  reading out its value.

We assume that all of the circuitry introduces errors to the device at a low rate. 
QEC proceeds by executing repeated rounds of stabilizer readout circuits to acquire a history of error syndromes, and then attempts to recover from the errors that may have occurred by using a decoder to interpret the syndrome data. 
The code distance is defined as the smallest number of error events that can lead a code to an undetected logical failure. 
The code distance can be increased by preparing a larger code consisting of more qubits. 
At low enough error rates one expects the logical failure rate of the code to improve as the distance increases.

\paragraph*{\it Identifying and mitigating defects.---} While early theoretical work on quantum error correction primarily considered error events as occurring uniformly on all qubits and components, extensive benchmarking on real devices reveals much more variation in the error rates across different qubits and couplers.  
Such variations can occur due to a range of possible error sources, but static variations in solid-state qubit processors can often be attributed to the fabrication process.  
Efficient benchmarking and error reconstruction techniques such as Refs.~\cite{Gambetta2012, carignandugas2023errorreconstructioncompiledcalibration, flammia2021b,hockings2024}, can be used to map out the landscape of qubit and component performance.  

One approach to accommodate such variability in the error profile, including the presence of defective components, is to operate the circuits for syndrome extraction as in the ideal case but to inform the decoder about the variation in device characteristics.  
With the surface code, state-of-the-art decoders such as BeliefMatching~\cite{higgott2023improved} or neural network decoders~\cite{Bausch2024,Varbonov2025,gu2026scalableneuraldecoderspractical} can use detailed information about the relevant error model of the device to identify the most likely errors in the decoding process.  
Such noise-informed decoding has been demonstrated to offer significant advantages in QEC, especially when the variability is small and does not give rise to any exceptionally poor or defective components~\cite{lee2026scalable,Google24}.  
Indeed, past successful demonstrations have been achieved by selecting only high-quality and spatially-uniform footprints on a device; while this approach can be used for near-term demonstrations, it limits the use of the full redundancy of a large device and can place unreasonable demands on fabrication yield.

Here we consider active strategies that modify the code and syndrome extraction circuits to avoid the most defective components.  
It is a subtle point to identify when a component should be flagged as defective or underperforming, thereby warranting active exclusion. 
In practice, we might expect the error rates of all the qubits of the device to respect some normal distribution about the mean where a small number of underperforming qubits lie in the tail of the distribution. 
For instance, for the operation of \texttt{ibm\_miami} in June 2026, IBM reported a median two-qubit gate error rate of $p\sim 2.83 \times 10^{-3}$ but the worst couplers had error rates up to $p \sim 1.69 \times 10^{-1}$.  
We identify any coupler with error rate above $5\%$ as underperforming, as well as couplers without any calibration data. 
Likewise, qubits have varying error rates. For measure qubits, the measurement error is the most relevant metric and we observe a median of $p \sim 2 \times 10^{-2}$; extreme cases may have measurement error rates as high as $1.5 \times 10^{-1}$. 
We consider measure qubits with error rates above $7 \times 10^{-2}$ as underperforming. For data qubits, we focus on $T_1$, the energy relaxation time over which an excited qubit decays to its ground state, and $T_2$, the dephasing time over which the qubit retains phase coherence. 
We observe a median $T_1 \sim 335\,\mu\mathrm{s}$ and a median $T_2 \sim 256\,\mu\mathrm{s}$, with extreme cases falling below $10\,\mu\mathrm{s}$ for both $T_1$ and $T_2$. 
We consider a data qubit defective when both $T_1$ and $T_2$ are under $80\,\mu\mathrm{s}$. In Fig.~\ref{fig:layout}, an example characterisation of the \texttt{ibm\_miami} device on a particular day (indicated above the device) identifies the underperforming components. 
We observe that it is not possible to place a standard distance-5 rotated surface code on the device using this geometry while avoiding defective components.

With a set of underperforming components identified, we must find the stabilizer group of a valid code that can be operated on a device in a way that avoids these components.
It is important that our stabilizer observables commute such that we can obtain all of their syndrome data simultaneously, thereby enabling us to precisely determine a correction that will reverse errors, up to the number of correctable errors, that may have occurred. 
This means that, if we find some underperforming components that we wish to exclude from our error-correction experiment, it is a non-trivial exercise to find a new code with a set of commuting stabilizers whose code distance is comparable to the largest distance that a perfect device could accommodate. 
Furthermore, we need to find an efficient readout schedule to learn the values of the stabilizers of the new code in a timely manner, otherwise we might expect a significant amount of noise to accumulate over the duration of the modified stabilizer readout circuit.

A central tool that enables us to avoid using underperforming components is commonly known as a `super-stabilizer'~\cite{Stace2009thresholds, Auger2017, Strikis2021, Siegel2023adaptivesurfacecode, Lin2024codesign, McLauchlan2024accommodating, Wei2025low, Leroux2025snakes, Debroy2025luciinsurfacecode,  wolanski2026automated, wei2026adaptivedeformationcolorcode,higgott2025handlingfabricationdefectshexgrid, mishmash2025excising}. 
When a component is excluded from the syndrome extraction circuit, neighboring stabilizers are merged into higher weight super-stabilizers surrounding the defect, allowing both Pauli-X and Pauli-Z errors in its vicinity to remain detectable. 
Rather than measuring these high-weight operators directly, their values are inferred from a collection of lower-weight checks whose product forms the corresponding super-stabilizer. 
While the resulting super-stabilizers commute, the lower-weight checks used to infer their values generally do not, so they must be measured in separate syndrome extraction rounds.

In our approach, isolated underperforming couplers and measure qubits are handled using the ``ancilla repurposing" method of Ref.~\cite{Leroux2025snakes}. Neighboring measure qubits are repurposed to perform additional parity checks, allowing the original stabilizers to be reconstructed while avoiding the defective components. For isolated coupler and measure qubit defects, this construction preserves the spatial code distance. Underperforming data qubits are accommodated using the super-stabilizer construction of Ref.~\cite{Auger2017, Strikis2021}, where neighboring stabilizers are merged around the excluded data qubit. This construction reduces the code distance in at least one Pauli basis. In the experiments below, we therefore distinguish between measure qubit/coupler defects and data qubit defects.

We illustrate the details of our surface-code experiments, together with a two-step measurement sequence that removes the need to use defective components, in Fig.~\ref{fig:layout}.  

\begin{figure*}[t]
\centering
\begin{tikzpicture}
  
    \node[inner sep=0pt] (a) at (0,0)
        {\includegraphics[width=0.7\textwidth]{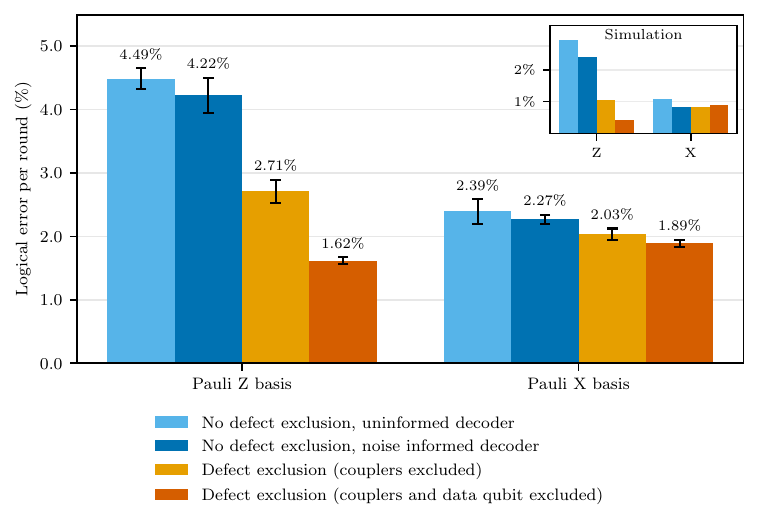}};
    \node[inner sep=0pt, anchor=west, yshift=1.4cm] (b) at ([xshift=0.4cm]a.east)
        {\includegraphics[width=0.3\textwidth]{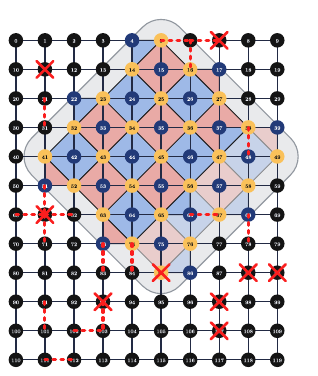}};
    \node[anchor=north west, inner sep=0pt]
        at ([xshift=4pt,yshift=5pt]a.north west) {(a)};
    \node[anchor=north west, inner sep=0pt]
        at ([xshift=2pt,yshift=5pt]b.north west |- a.north west) {(b)\hspace{1em}2026-05-11};

\end{tikzpicture}

    \caption{\textbf{Memory experiments for a distance-5 surface code.} (a)~Results of memory experiments performed in both the Pauli-Z and Pauli-X basis, comparing different strategies to accommodate defects. We perform sequences of error correction up to 32 rounds, and report logical error per round.  We compare four strategies.  First, as a simple baseline, the code and syndrome extraction circuit are unmodified, using an uninformed decoder where all the priors are chosen uniformly based on the median Pauli error rates from the calibration data.  Second, we consider a passive strategy, where the code and syndrome extraction circuits are left unmodified, and using a noise-informed decoding approach based the full calibration data; see Ref.~\cite{lee2026scalable}.  Third, we employ an active defect exclusion strategy where we avoid defective couplers using methods adapted from Ref.~\cite{Leroux2025snakes}. Finally, we apply a similar strategy where we also exclude a data qubit by modifying the circuit, removing an entire row in the surface-code layout, with a Pauli-X distance reduced to 4 (see (b)). The logical error per round is obtained by fitting the measured logical error probability $P_L$ as a function of the number of measurement rounds $r$ (from $r=$ 2 to 32) to $P_L(r)=\frac{1}{2}(1-a\,e^{-r/\tau})$, and reporting $\varepsilon=\frac{1}{2}(1-e^{-1/\tau})$, with uncertainty propagated from the fitted decay constant $\tau$.  The inset shows the results of simulations of these same experiments, performed in Stim, with circuit-level noise chosen according to the calibration data. (b) Data is acquired for a distance-5 surface code as shown, with defects as indicated for the device operating on the date shown.  The faded region indicates the row of the original code footprint that is removed for the data-qubit exclusion strategy.
    }
    \label{fig:memory_experiments_d3}
\end{figure*}

\paragraph*{\it Memory experiments.---} Figure~\ref{fig:memory_experiments_d3} presents the results of memory experiments using different strategies to accommodate defective components.  We find significant improvements in the logical error per round of our memory experiment when we employ active defect exclusion strategies. When we exclude underperforming couplers, we observe a logical error per round of $2.71\%$ for a memory experiment performed in the Pauli-Z basis, which is a significant improvement compared with the logical error per round of $4.22\%$ using only a decoder informed strategy, as we describe below. The improvement in the same memory experiment performed in the Pauli-X basis is more modest. 

Several factors may contribute to the asymmetry observed in our memory experiments performed in the two Pauli bases.  First, the defective couplers in our memory experiment participate only in Pauli-Z stabilizer measurements, which detect X-type errors. These errors can lead to logical-X errors and therefore affect the Z-memory performance more strongly. Additionally, the defective components are concentrated along a row supporting minimum-weight X logical operators, which may further increase the probability of Z-memory failures. Furthermore, our strategy for accommodating this arrangement of defects requires forming a large X-type super-stabilizer of weight-10, see Fig.~\ref{fig:layout}, and this can lead to reduced X-memory performance.
 
Even after excluding the defective couplers, a residual asymmetry remains between the memory experiments performed in the Pauli-Z and Pauli-X bases ($2.71\%$ vs $2.03\%$).
A possible cause for this residual asymmetry is that the measure qubits and couplers participating in the Z-stabilizer checks had higher mean readout ($2.61\%$ vs $2.43\%$) and two-qubit CZ ($0.40\%$ vs $0.29\%$) error rates than their X-check counterparts at the time of the experiment, which likely results in less reliable Z-syndrome extraction.

We observe a further reduction in logical error per round, to $1.62\%$ in the Pauli-Z basis, when we additionally exclude an underperforming data qubit. When excluding only the data qubit and not the couplers we do not observe a reduction in the logical error probability; see Appendix~\ref{app:data-only}. 
Again the improvements to the logical error per round in the Pauli-X basis are modest, reducing to $1.89\%$. 
Excluding this data qubit together with the surrounding underperforming components requires removing an entire row of the surface-code layout; see Fig.~\ref{fig:memory_experiments_d3} for details. 
The reduction in logical error probability in the Z basis is then partly attributable both to the data qubit defect removal, and to the smaller number of minimum-weight X-type logical strings, that can flip the Z logical, in this smaller instance. 
In this fully defect-removed code, the measured logical errors per round below $2\%$ in both bases are approaching, but not beating, the best distance-3 surface code; see Appendix~\ref{app:additional-d3}.

As a control experiment, we also exclude a measure qubit and coupler that are not identified as underperforming; see Appendix~\ref{good_comp_excluded}. We do not observe an improvement in logical error per round in this case.

We compare active defect exclusion strategies against a passive strategy where the code and syndrome extraction circuit are left unmodified and we use a noise-informed belief-matching decoder with a detector error model built from the calibration data, see Ref.~\cite{lee2026scalable}. We also compare against an uninformed decoder where all the priors are chosen uniformly, using the median Pauli error rates from the calibration data. We observe a much more marginal improvement using noise-informed decoding, e.g., reducing the logic error rate per round in the Pauli-Z basis from $4.49\%$ using the uninformed decoder to $4.22\%$ with the noise-informed decoder, compared to active defect exclusion strategies. 

For all experiments reported here, decoding is performed using BeliefMatching~\cite{higgott2023improved}. Unless stated otherwise, we furnish the BeliefMatching decoder with a detailed error model informed by IBM calibration data as described in Appendix~\ref{calib_data}. The construction of the detection events used for decoding is described in Appendix~\ref{app:no-reset}. We have simulated all of our strategies for accommodating defects in these memory experiments, using Stim \cite{gidney2021stim}, with a noise model obtained from the calibration data on the day of the experiment.  We observe qualitatively similar behaviour as the experiments, but consistently lower error rates. We attribute this discrepancy in part to noise sources that are not captured with our circuit-level noise model, such as leakage.

\paragraph*{\it Stability experiments.---}In addition to parity measurements being necessary for quantum error correction of a memory circuit, they also play a key role in performing error-corrected measurement-based logic gates in fault-tolerant quantum computing. For instance, in lattice surgery~\cite{Horsman2012,fowler2012b,Litinski2019gameofsurfacecodes,fowler2019lowoverheadquantumcomputation}, we perform a set of local checks, the product of which reads out an observable of some logical degree of freedom. Given that we need to reliably learn the value of a large subset of checks, we must identify measurement errors that may occur. In order to identify measurement errors, we repeat these checks multiple times, and decode across the entire syndrome history. We can decrease the logical error probability by increasing the number of times we repeat these checks, assuming we are below some threshold rate of error.

\begin{figure*}[t]
\centering
\begin{tikzpicture}
  
    \node[inner sep=0pt] (a) at (0,0)
        {\includegraphics[width=0.7\textwidth]{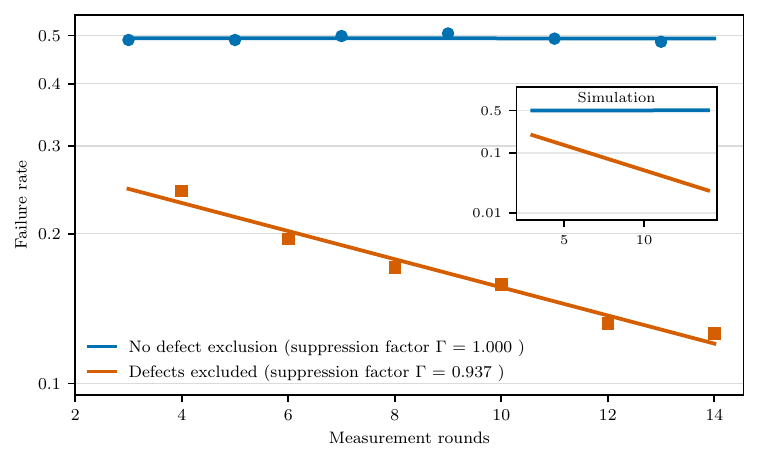}};
    \node[inner sep=0pt, anchor=west, yshift=0.5cm] (b) at ([xshift=0.4cm]a.east)
        {\includegraphics[width=0.29\textwidth]{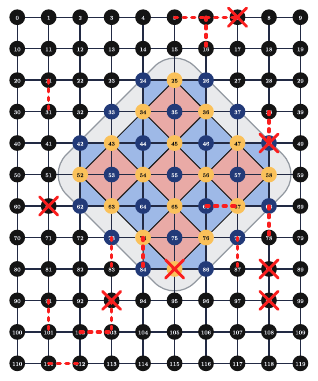}};
    \node[anchor=north west, inner sep=0pt]
        at ([xshift=4pt,yshift=5pt]a.north west) {(a)};
    \node[anchor=north west, inner sep=0pt]
        at ([xshift=2pt,yshift=5pt]b.north west |- a.north west) {(b)\hspace{1em}2026-06-17};

\end{tikzpicture}

    \caption{{\bf Stability experiments.}  (a) Failure rate of the stability experiment as a function of the number of measurement rounds for two strategies. Blue: a strategy in which the syndrome measurements are left unmodified and noise-informed decoding is performed using the full calibration model. Orange: a strategy in which defective couplers and measure qubits are excluded, following the approach shown in Fig.~\ref{fig:layout}(e--f). In this experiment, the defective data qubit within the code is retained and used as a regular data qubit. The failure rate is fit to the exponential form $F = C\Gamma^{r}$, where $C$ is an unknown prefactor and $r$ is the number of measurement rounds. The fitted values of $C$ are $C=0.49$ for the unmodified strategy and $C=0.30$ for the defect exclusion strategy. The inset shows simulations of these experiments, performed in Stim, using a circuit noise model based on calibration data. (b) Layout of the stability experiment on the device, with defects as indicated for the device operating on the date shown. A corresponding stability experiment performed on a code without a defective data qubit is shown in Fig.~\ref{fig:Miami_stability2}; it yields a comparable fitted value of $\Gamma$, suggesting that the contribution of the defective data qubit to the observed failure rate is small.}
    
    \label{fig:Miami_stability}
\end{figure*}

\begin{figure*}[t]
\centering
\begin{tikzpicture}
  
    \node[inner sep=0pt] (a) at (0,0)
        {\includegraphics[width=0.7\textwidth]{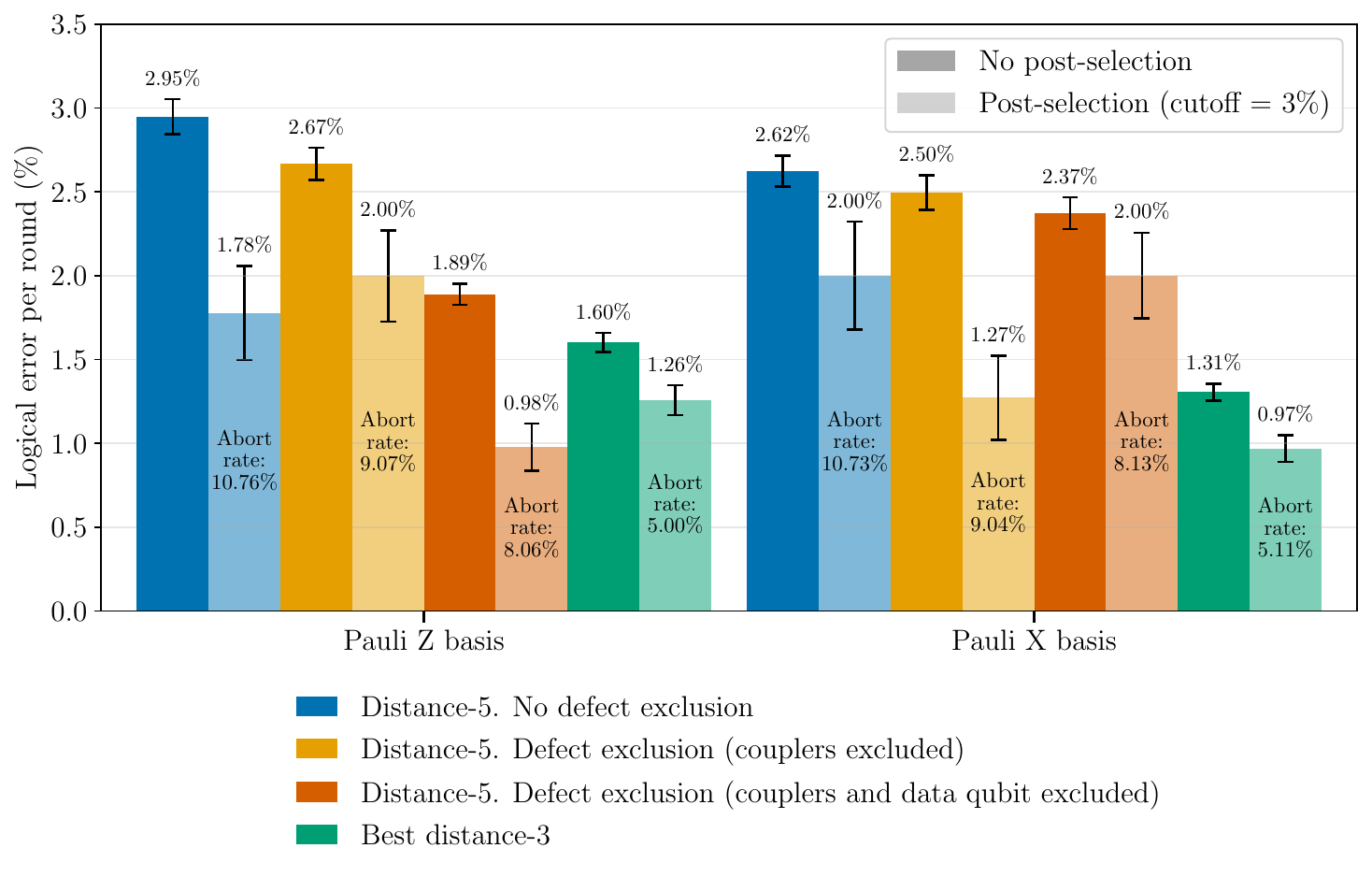}};
    \node[inner sep=0pt, anchor=west, yshift=0.8cm] (b) at ([xshift=0.4cm]a.east)
        {\includegraphics[width=0.3\textwidth]{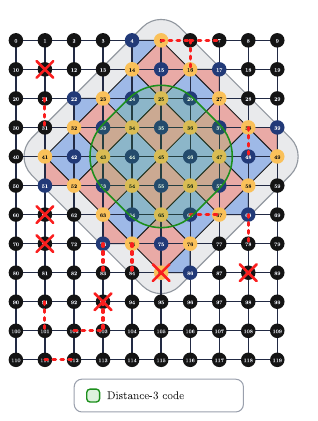}};
    \node[anchor=north west, inner sep=0pt]
        at ([xshift=1pt,yshift=8pt]a.north west) {(a)};
    \node[anchor=north west, inner sep=0pt]
        at ([xshift=2pt,yshift=8pt]b.north west |- a.north west) {(b)\hspace{1em}2026-06-23};

\end{tikzpicture}

    \caption{{\bf Post-selecting leakage in memory experiments.} (a)~Logical errors per round of memory experiments with leakage post-selection in different bases. Three different strategies for accommodating defects are shown, where logical error per round with and without leakage post-selection are shown side by side. For comparison, we also perform a memory experiment on a distance-3 code, where we determine the best instance for this distance-3 code within the footprint of the distance-5 code using circuit-level simulations of the device prior to the experiment. This distance-3 code uses no defective components. 
    (b)~Layout of the distance-5 surface code on the device, with defects as indicated for the device operating on the date shown.  The distance-3 surface code footprint is also indicated in green overlay. 
    \label{fig:leakage_post_selection}}
\end{figure*}

Measurement-based logic gates are especially sensitive to underperforming components, since such components can lead to errors across a sequence of time steps, which may build up to a logical readout error. As such, we should also consider modifying logic gates to quarantine defective components to improve the performance of these logic gates. Once again, we can build super-stabilizers and use other modifications to infer our logical observable while avoiding substandard components. Given that we should expect underperforming qubits to affect measurement-based logic gates differently from a memory experiment, we need to separately test the effect of defect mitigation on logical operations. 

To this end we perform stability experiments~\cite{Gidney2022dual, Geher:2025aa, harper2025characterising} to investigate the effect of defective components on measurement-based logic gates. A stability experiment is designed such that the product of stabilizer checks of a given Pauli type produces a known value, usually $+1$. Testing how reliably we obtain this known value after decoding serves as a useful proxy for the reliability of operating a logic gate on the hardware. 

We improve our ability to detect measurement errors in stability experiments by repeating our check measurements multiple times. If we perform $r$ rounds of syndrome measurement, these repeated checks should lead to a reduction in our failure rate $F$; the probability of inferring the wrong value for the product of stabilizer checks after error correction. This quantity is expected to behave as $F = C\Gamma^r$  for a suppression factor $\Gamma$ and some constant $C$ that depends on the stability experiment details.  A suppression factor $\Gamma <1$ indicates that repeated rounds of syndrome measurement lead to an improvement. Alternatively, $\Gamma \ge 1$ indicates that the error correction strategy is introducing errors at a faster rate than it is able to correct them, and repeated rounds of syndrome measurement are not helping. We use $\Gamma$ to quantify the performance of the stability experiment.

Details of our stability experiment are illustrated in Fig.~\ref{fig:layout}. 
The boundaries are defined such that the product of Pauli-Z checks gives the identity. We define a successful sample as one in which the decoder correctly identifies the number of measurement errors (modulo 2) over the subset of checks that give identity. We compare the following two cases.  In the benchmark case, we do not modify the syndrome extraction circuits of our stability experiment, and inform the decoder with calibration data.  In the case of our defect exclusion strategy, we construct a code and syndrome extraction circuit that avoids defective components, as shown in Fig.~\ref{fig:layout}.

Results of the stability experiments are presented in Fig.~\ref{fig:Miami_stability}. We find that we can reliably infer the value of the stability observable only when avoiding underperforming components. When an active defect exclusion strategy is used, we observe that the failure rate decays as we increase the number of rounds of syndrome extraction, with a measured suppression factor $\Gamma = 0.94$.  In contrast, without such a strategy, we do not observe a decay in the failure rate as the number of measurement rounds is increased; $\Gamma \geq 1$. We note that, in our defect exclusion strategy, some checks are measured less frequently -- once every two rounds of syndrome extraction rather than once per round -- thereby reducing the circuit-distance of the stability experiment. This means that we need to complete more rounds of syndrome extraction to obtain a commensurate distance. We plot our graph according to `measurement rounds', where we count the number of rounds of mid-circuit measurements we make over the experiment, noting that a full syndrome extraction cycle is completed only every second measurement round.  

We have simulated our stability experiments, using Stim, with these same strategies and using the calibration data to construct a Pauli noise model. 
Our simulation results capture the qualitative behaviour observed in experiment, but for the active defect exclusion strategy the simulations predict a larger suppression factor of $0.83$, suggesting, as with the memory experiments, that the error model used in simulations does not capture all error processes.

\paragraph*{\it Leakage.---}We observe significant improvements in QEC performance by using strategies that exclude defects.
However, while our simulations, shown in figure insets, based on the measured calibration data capture these improvements qualitatively, they indicate the presence of additional error processes in the experiments that are not captured in this calibration error model.  
Leakage out of the qubit subspace is one likely source of such errors.  
While the capabilities of \texttt{ibm{\_}miami} do not currently include a mid-circuit `reset' or related method to restore a transmon to its computational subspace, we can quantify the impact of leakage on logical error probabilities using post-selection on the IQ data returned from a measurement, following Ref.~\cite{lee2026scalable}. 
For each measurement we perform, we take the IQ plane data, which enables us to estimate the probability that the transmon has escaped from its computational subspace. 
We then post-select on the set of samples where the likelihood of a leakage error having occurred is below a certain cutoff.

Memory experiments using various strategies to accommodate defective components combined with leakage post-selection are shown in Fig.~\ref{fig:leakage_post_selection}.  
We consistently observe a significant reduction in the logical error per round when we post-select on samples where it is unlikely that leakage has occurred. 
Once again, across various strategies, the most substantial improvements are observed in memory experiments performed in the Pauli-Z basis. 
We compare our results to the best distance-3 code within the footprint of our distance-5 experiment, where we determine the best distance-3 code using simulations in advance of the experiment. 
The distance-3 patch used is free of any defective components. 
We find for memory experiments in the Pauli-Z basis that, after leakage post-selection with a cutoff leakage probability of $3\%$, using circuit-level defect exclusion for all of our defective components, our distance-5 code has a logical error per round of $0.98\%$ whereas the distance-3 code has $1.26\%$. 
We do not observe such a similar improvement in memory experiments performed in the Pauli-X basis, and we attribute this to the fact that removing the data qubit reduces the distance associated with Pauli-Z errors.

\paragraph*{Discussion.---}We have investigated strategies for quantum error correction, and error-corrected logic gates, on a large quantum processor with a handful of underperforming components. 
We find performance consistently improves by a significant margin using active defect exclusion strategies. 
These tools will become essential as we continue to produce larger devices that will inevitably include a small fraction of underperforming or defective components.

Our stability experiment results in particular demonstrate that defect mitigation is crucial for performing error-corrected, measurement-based logic gates. 
When defects are removed, we observe not only a significant level of per-round error suppression that was not previously present, but also an overall decrease in the failure rate, even at low numbers of measurement rounds. 
We therefore see evidence for below-stability-threshold behaviour, where increasing the number of measurement rounds suppresses logical failure rates, in the cases where defects are removed.

A subtle point that is deserved of further investigation is how we define a defect. 
It is common in the theory literature to model a defect as a qubit that is not addressable by any means. 
In practice, on this device, many of the defective components are functional, but they operate at an error rate several times higher than the average. 
It is worthwhile investigating how the probability of a logical error is affected by redefining how we identify and exclude underperforming components. 
It may well be that further improvements could be made by being more selective of the components we use. Of course, if we are too selective, we do not have enough components left to operate an error-correction experiment. 
By exploring this tradeoff we might find further improvements.

We observed that postselection of events based on the probability of no leakage occurring led to significant reductions in logical error probabilities for our memory experiments across all strategies; however, while this postselection can provide insight into the effects of leakage, it does not provide a scalable approach to improving QEC.
A scalable approach would require additional device capabilities, such as the ability to reset into the qubit subspace together with modified syndrome extraction circuits that enable us to continually reset and refresh qubits as we read out stabilizer data~\cite{Fowler2013, Suchara2014, Magnard2018, Miao2023, eickbusch2025demonstratingdynamicsurfacecodes, Google24}. 
Our experiments suggest it will be important to combine leakage-resilient syndrome extraction circuits with strategies for active exclusion of defects.  These results highlight the critical role of co-designing error-correcting codes and logical gate implementations with the underlying noise characteristics of the hardware, particularly as superconducting quantum processors are scaled to larger sizes.

\section*{Data availability}
The data generated in this study have been deposited in the Zenodo 
database~\cite{datarep}. %\cite{Zenodo}

\begin{acknowledgements}
We are grateful for conversations with E.~Pritchett, S.~Singh, K.~Siva and M.~Takita.  B.J.B.~is grateful for the hospitality of the Center for Quantum Devices at the University of Copenhagen. We acknowledge support from the Intelligence Advanced Research Projects Activity (IARPA), under the Entangled Logical Qubits program through Cooperative Agreement Number W911NF-23-2-0223 (R.H., B.J.B., and S.D.B.).  The views and conclusions contained in this document are those of the authors and should not be interpreted as representing the official policies, either expressed or simplied, of IARPA, the Army Research Office, or the U.S. Government. The U.S. Government is authorized to reproduce and distribute reprints for Government purposes notwithstanding any copyright notation herein. S.H.L.~was supported by the National Research Foundation of Korea (NRF) grant (RS-2026-25476454, RS-2024-00442710, RS-2023-NR068116) funded by the Ministry of Science and ICT (MSIT) of the Korean government. 
D.J.W.~is supported by the Australian Research Council Discovery Early Career Research Award (DE220100625).
\end{acknowledgements}


%\newpage
\appendix
\section{Mid-circuit measurements}

\begin{figure*}[t]
\begin{tikzpicture}
\node[inner sep=0pt] at (-4,0)         {\includegraphics[width=0.5\textwidth]{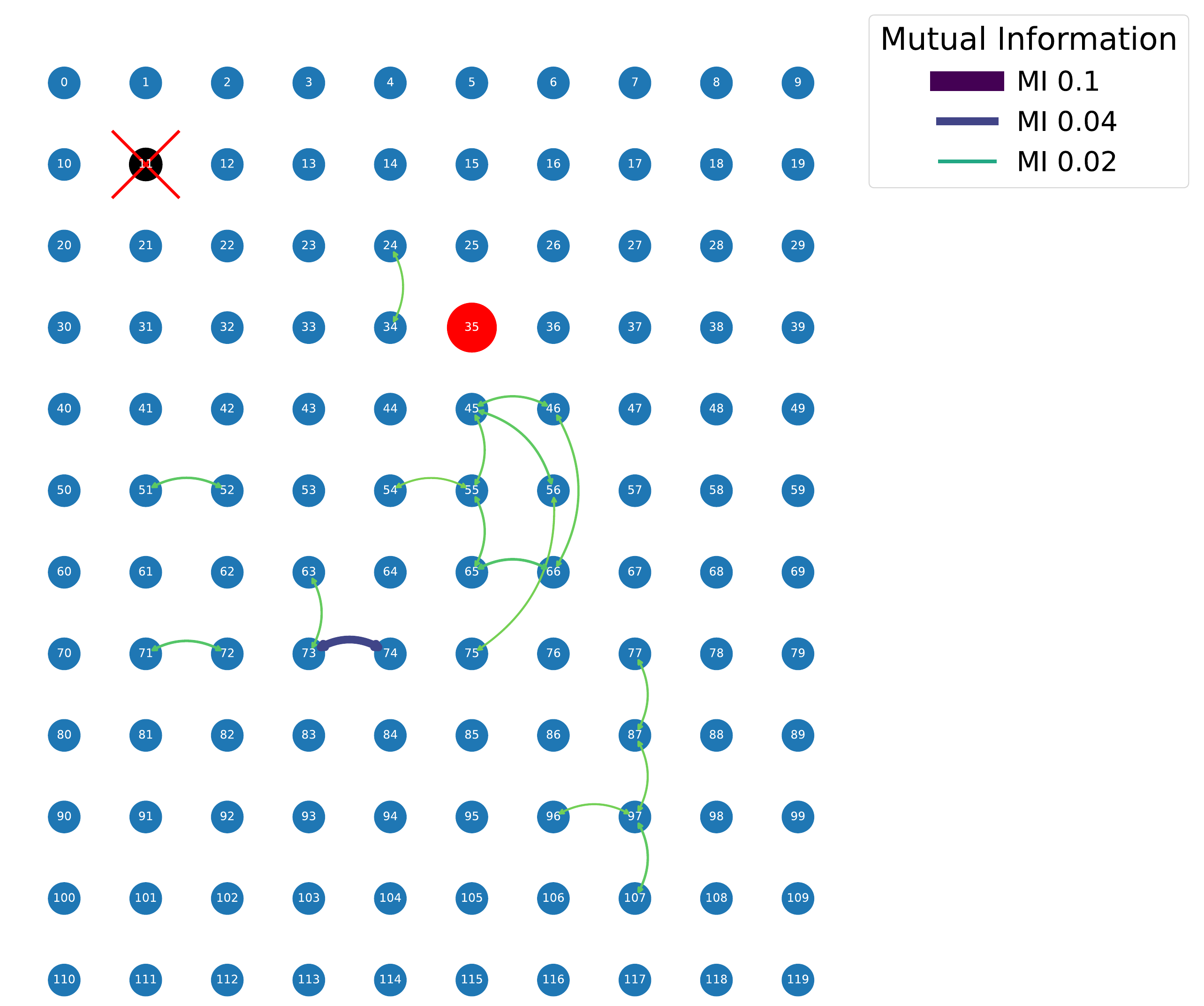}};
\node[inner sep=0pt] at (5,0)         {\includegraphics[width=0.5\textwidth]{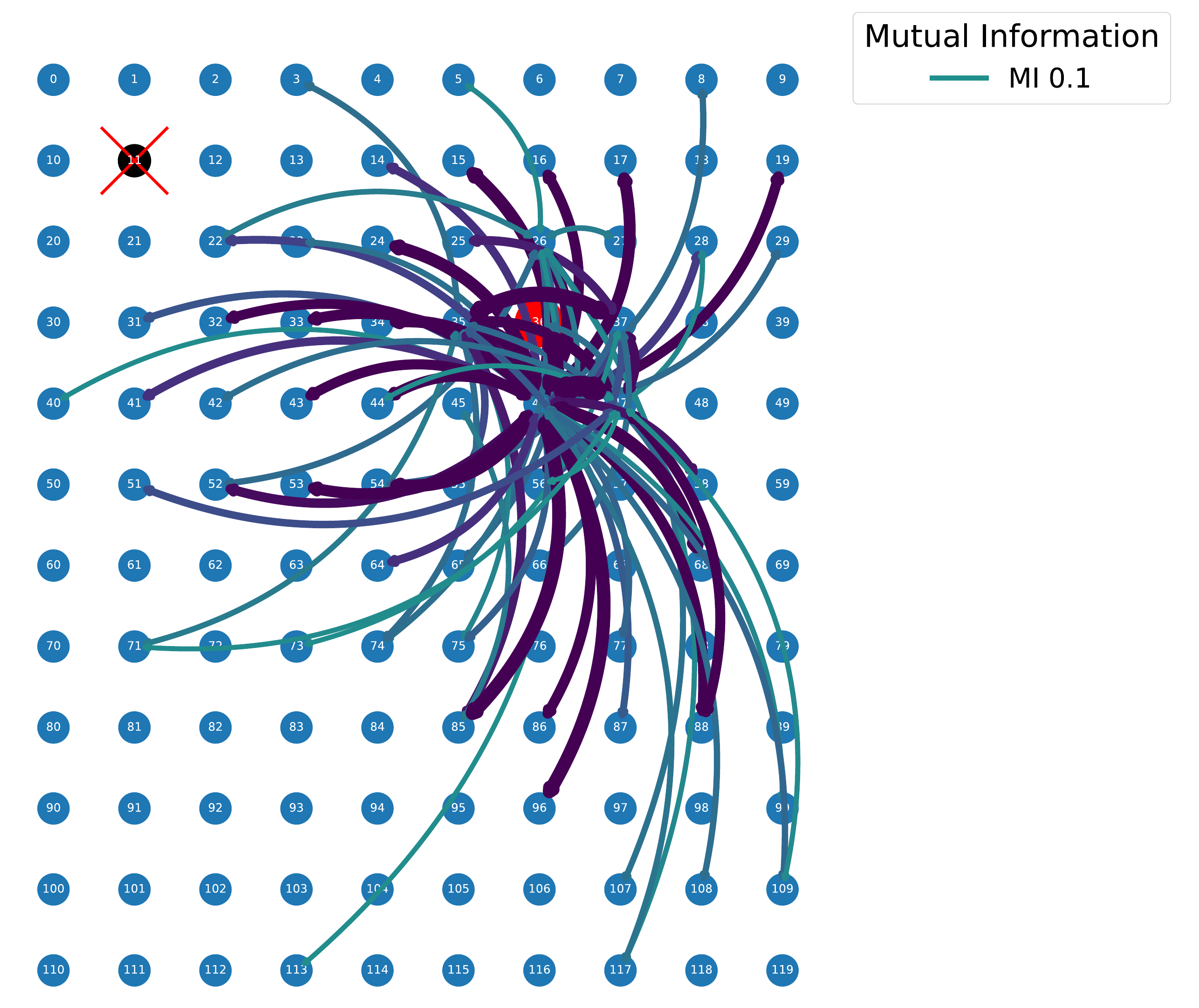}};
\node[] at (-8,4) {(a)};
\node[] at (1,4) {(b)};
\end{tikzpicture}	
\centering
    \caption{\textbf{Correlations due to mid-circuit measurements.} \textbf{(a)} Mutual information between the qubits during a simultaneous single qubit Clifford twirl. The qubit in red (here qubit 35) is measured every 4th timestep during the protocol (see text for more details). Qubit 11 has been marginalised out because of a high error rate. The lines joining the qubits show those qubits that exhibit a mutual information above a 0.01 (1 bit in 100) threshold, the color and width of the line indicating the strength of the mutual information.  For this choice of special qubit 35, all mutual information between qubits is relatively modest.
    %While some mild mutual information is present, if this device is similar to the Heron devices this likely follows readout resonator patterns. 
    \textbf{(b)} Mutual information between qubits, using an identical protocol to (a), but with the mid-circuit measurement being performed on qubit 36 (as opposed to qubit 35). Here the cut-off for display is 0.1 (1 bit in 10) and as can be seen the effect of performing mid-circuit measurements on this qubit leads to extremely high levels of noise on many other qubits. Note that in order to prevent the line widths overwhelming the figure, the scale is different from the graph in~(a).  }\label{fig:mi}
\end{figure*}

\begin{figure}[t]
    \centering
    \includegraphics[width=0.45\textwidth]{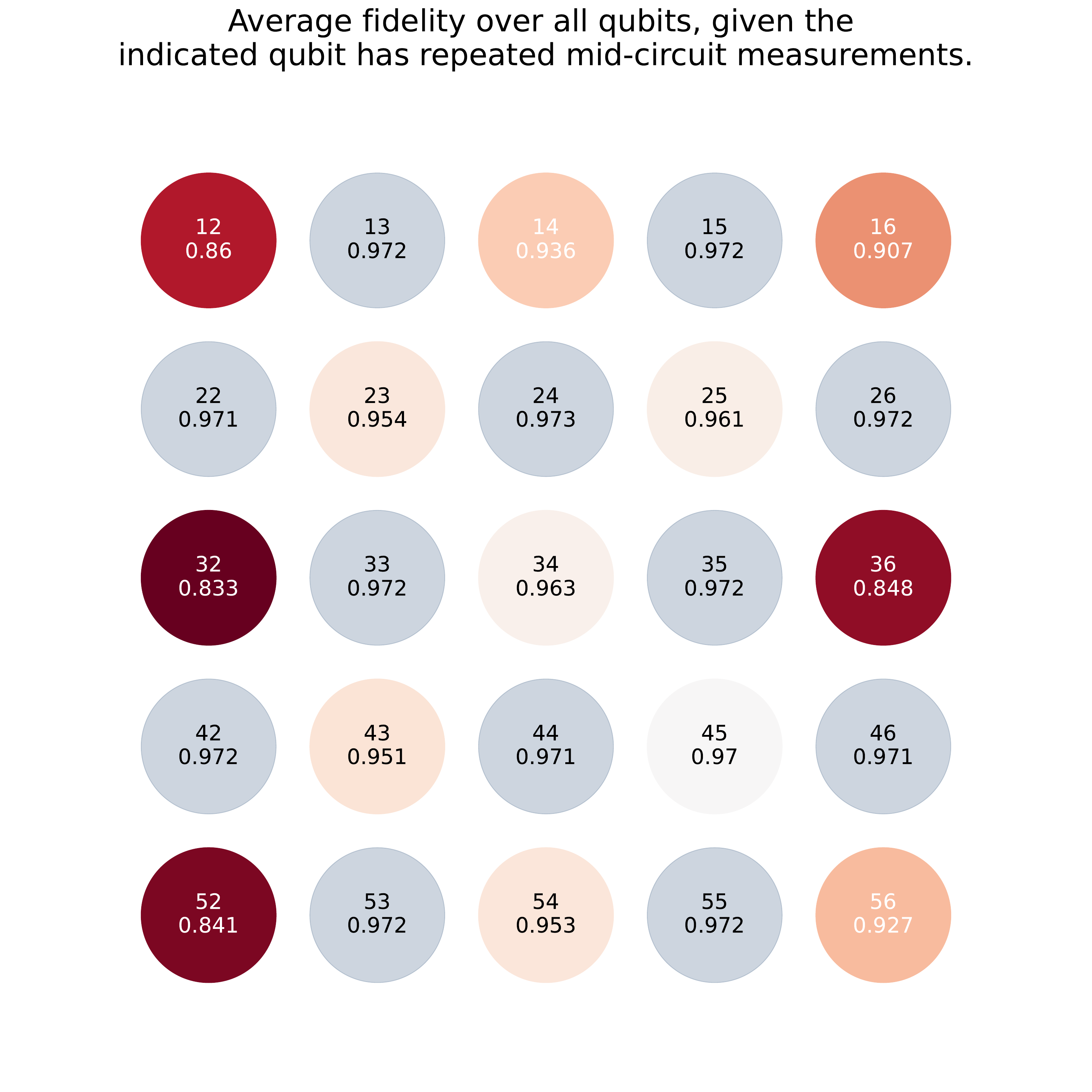}
    \caption{\textbf{Impact of mid-circuit measurements.} Using our approach to characterise the effect of specific mid-circuit measurements on spectator qubits, as described in the text and \Cref{fig:mi}, we map the impact of noise correlated with mid-circuit measurements on a $5\times 5$ grid of qubits on the quantum processor. The characterisation protocol was performed for each choice of 25 locations for the special qubit, and the average fidelity of all \emph{other} qubits in the device is reported. Reemphasising this point, the fidelities reported here do not correspond to the fidelity of the qubit, but rather the average fidelities of all the qubits if the indicated special qubit is subjected to a mid-circuit measurement. In the absence of cross-talk, we would expect the average fidelity to remain relatively unchanged, regardless of which qubit is repeatedly measured. This is the case for all of the grey coloured qubits, where the fidelities are the same $\pm 0.001$, within shot noise. For other choices of special qubit, measuring that qubit can reduce the fidelity of the other qubits in the device. This is indicative of the cross-talk mechanism illustrated in \Cref{fig:mi}.   }\label{fig:fidelities}
\end{figure}

The quantum processor \texttt{ibm\_miami} is classified as an exploratory QPU and in particular its release notes specified that, ``the quality of mid-circuit measurements is limited.'' In order to understand the impact of the mid-circuit measurements on our codes, we used a simple benchmarking protocol loosely based on Refs.~\cite{Govia2023,Beale2023}. Simultaneous single Clifford-twirl benchmarking was performed on all of the qubits with a modification applied to one specific qubit.  For this special qubit, every four gate timesteps, it is returned to the computational basis and measured.  During this measurement, XY4 dynamic decoupling \cite{Viola1999,Gonzalo_2010} is performed on the other qubits. Using the analysis detailed in Refs.~\cite{Harper2019b,Harper2023,harper2025characterising}, we extract the joint probability distribution between arbitrary qubits and calculate the mutual information between those qubits. Since only single qubit operations are being performed in the ideal case, there should be zero mutual information between the qubits. \Cref{fig:mi} shows two specific examples where mid-circuit measurements are performed on qubit 35 or qubit 36 as the special qubit. There is a marked disparity between the quality, not only of the measurement but also on the noise affecting many spectator qubits. While we did not check every qubit on the device, \Cref{fig:fidelities} shows the result of a scan of a $5 \times 5$ block of 25 qubits. From these data, if the digits of a qubit sum to an even number, mid-circuit measurements performed on that qubit do not induce a significant amount of noise on other qubits, whereas if they sum to an odd number, the impact of mid-circuit measurements led to excess noise on the other qubits in the device.  While this has an impact on the placement of error correcting codes, for our purposes we just ensured that all measure qubits in the code were `even' qubits, allowing us to obtain the results presented in the body of this paper.

\section{Additional experiments}
\subsection{Effect of excluding only the defective data qubit} \label{app:data-only}
\begin{figure*}[t]
\centering
\begin{tikzpicture}
  
    \node[inner sep=0pt] (a) at (0,0)
        {\includegraphics[width=0.7\textwidth]{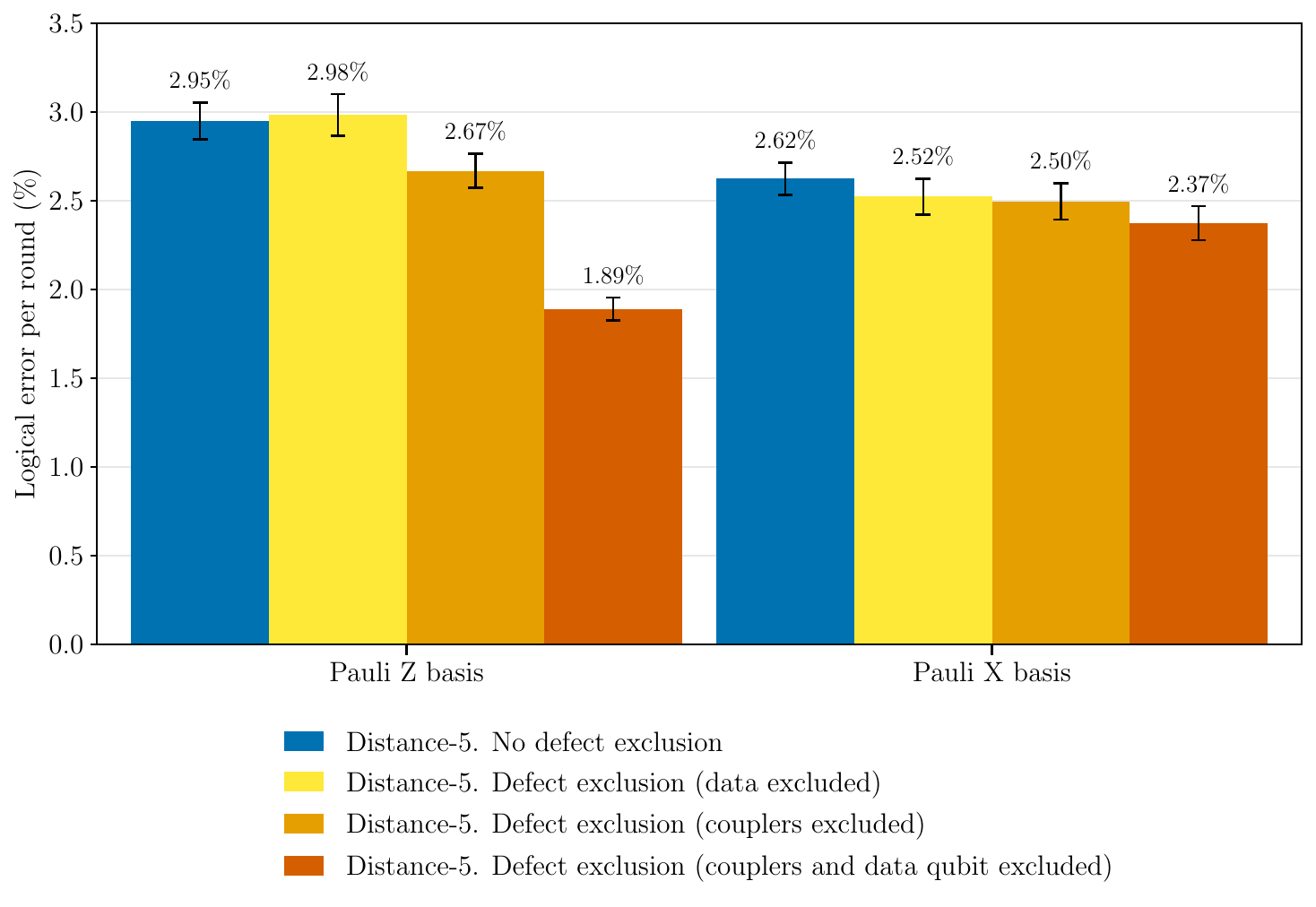}};
    \node[inner sep=0pt, anchor=west, yshift=1.4cm] (b) at ([xshift=0.4cm]a.east)
        {\includegraphics[width=0.3\textwidth]{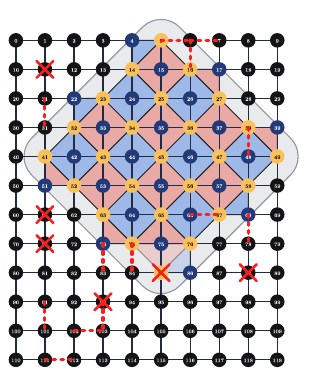}};
    \node[anchor=north west, inner sep=0pt]
        at ([xshift=1pt,yshift=7pt]a.north west) {(a)};
    \node[anchor=north west, inner sep=0pt]
        at ([xshift=2pt,yshift=7pt]b.north west |- a.north west) {(b)\hspace{1em}2026-06-23};

\end{tikzpicture}

    \caption{(a) Effect of removing each defective component separately. This is the same data as in \cref{fig:leakage_post_selection} with no post-selection, but with the addition of a circuit that only excludes the data qubit. Removing only the data qubit does not improve the logical error per round in both bases. (b) Data is acquired for a distance-5 surface code as shown, with defects as indicated for the device operating on the date shown.  The faded region indicates the part of the original code footprint that is removed for the only data-qubit exclusion strategy.}
    \label{fig:no-data}
\end{figure*}

The memory experiment presented in Fig.~\ref{fig:memory_experiments_d3} yields the best results when an underperforming data qubit is excluded in addition to underperforming measure qubits and couplers. To investigate whether this improvement is due simply to the removal of the underperforming data qubit, we performed an additional memory experiment where only this data qubit was excluded while keeping the rest of the syndrome extraction circuit unchanged. The results are shown in Fig.~\ref{fig:no-data}. In contrast to the strategy where defective measure qubits and couplers are excluded, this modification does not result in a significant improvement in the logical error per round, while maintaining the same footprint on the device. This indicates that simply removing the defective data qubit is not sufficient to mitigate the dominant error sources.

In the experiments from Fig.~\ref{fig:memory_experiments_d3}, the exclusion of the defective data qubit is therefore combined with the removal of surrounding defective measure qubits and couplers. Due to the clustering of defects on the device, this requires removing an entire row of the surface code, resulting in a reduced $5\times4$ surface-code with reduced Pauli-$X$ distance, as described in Fig.~\ref{fig:memory_experiments_d3}.

\subsection{Additional distance-3 memory experiment} \label{app:additional-d3}
\begin{figure*}[t]
\centering
\begin{tikzpicture}
  
    \node[inner sep=0pt] (a) at (0,0)
        {\includegraphics[width=0.55\textwidth]{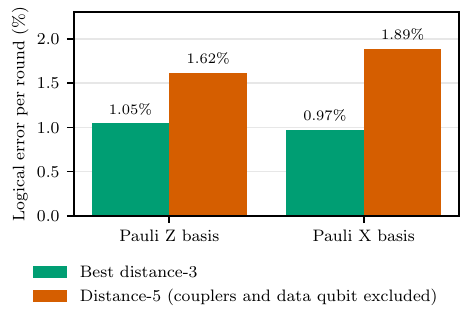}};
    \node[inner sep=0pt, anchor=west, yshift=0.3cm] (b) at ([xshift=0.4cm]a.east)
        {\includegraphics[width=0.3\textwidth]{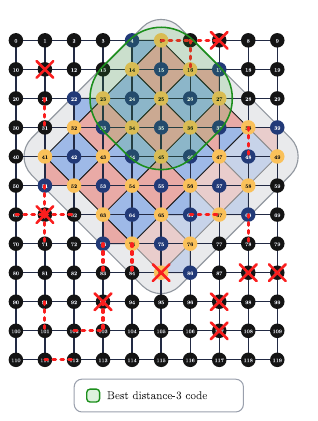}};
    \node[anchor=north west, inner sep=0pt]
        at ([xshift=4pt,yshift=10pt]a.north west) {(a)};
    \node[anchor=north west, inner sep=0pt]
        at ([xshift=2pt,yshift=10pt]b.north west |- a.north west) {(b)\hspace{1em}2026-05-11};

\end{tikzpicture}

    \caption{(a) Memory experiment results for the best distance-3 surface code inside the distance-5 surface code presented in \cref{fig:memory_experiments_d3}. (b) Surface code instance used to acquire the data in (a). Note that the distance 3 code does not contains any underperforming components}
    \label{fig:bestd3}
\end{figure*}

In \cref{fig:bestd3}, we show the results of a memory experiment on the best distance-3 patch, taken at the same time as the data in \cref{fig:memory_experiments_d3}. 
We find that the distance-3 code performs better than the the best defect exclusion circuit for the distance-5 patch in \cref{fig:memory_experiments_d3}. The distance-3 surface code instance does not use any defective components.
We did not take IQ data on this day and therefore cannot perform leakage post-selection like we did in \cref{fig:leakage_post_selection}.
However, as this was the best performing distance-3 code that we observed on any date, we expect that the gains from leakage post-selection for the distance-3 code would be less than in \cref{fig:leakage_post_selection}.

\subsection{Excluding good components}\label{good_comp_excluded}

\begin{figure*}[t]
\centering
\begin{tikzpicture}
  
    \node[inner sep=0pt] (a) at (0,0)
        {\includegraphics[width=0.7\textwidth]{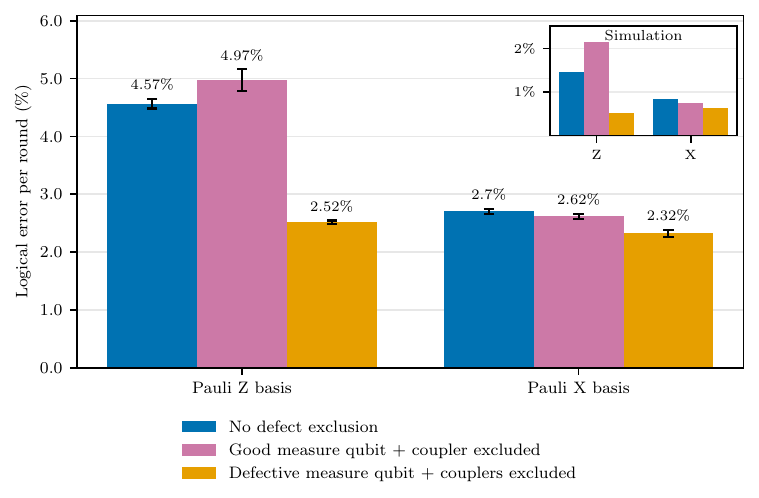}};
    \node[inner sep=0pt, anchor=west, yshift=0.7cm] (b) at ([xshift=0.4cm]a.east)
        {\includegraphics[width=0.3\textwidth]{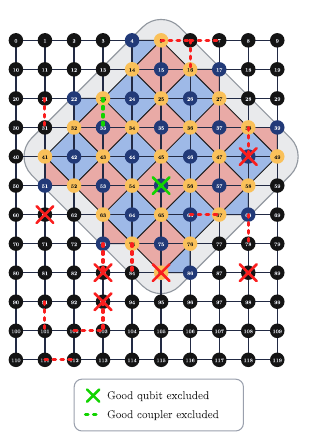}};
    \node[anchor=north west, inner sep=0pt]
        at ([xshift=4pt,yshift=5pt]a.north west) {(a)};
    \node[anchor=north west, inner sep=0pt]
        at ([xshift=2pt,yshift=5pt]b.north west |- a.north west) {(b)\hspace{1em}2026-06-10};

\end{tikzpicture}

    \caption{{\bf Excluding different couplers.} (a)~Logical error per round for memory experiments on a distance-5 surface code with different choices of couplers removed. We compare three cases.  First, we apply a strategy where the code and syndrome extraction circuit are unchanged, with decoding informed by the full calibration data.  Second, we apply our exclusion strategy by removing an ordinary qubit and coupler that are not underperforming.  Third, we exclude only the defective measure qubit and couplers. We note that all of the excluded couplers are associated to the Pauli-Z stabilizers and as such the improvements are more marked in the Pauli-Z basis.  (Inset (a)).~Simulations of this set of experiments, using Stim and the full calibration data as the noise model. (b)~Layout of the distance-5 surface code on the device, with defects as indicated for the device operating on the date shown.  }
    \label{fig:memory_experiments_d5}
\end{figure*}

To give more confidence in our method, in a second set of memory experiments, we compared our results to a control experiment where we treat a typical coupler as defective. These results are presented in Fig.~\ref{fig:memory_experiments_d5}. As expected, we observe that the removal of a typical coupler makes the experiment marginally worse, whereas removing defective couplers makes the Pauli-Z basis considerably better. The effects to the Pauli-X basis results are marginal because we only remove couplers associated with Pauli-Z stabilizer measurements. These results were collected on a separate run from those in Fig.~\ref{fig:memory_experiments_d3}, on a different date as indicated in the figure, and demonstrate a different set of underperforming components. 

\subsection{Additional stability experiment}
\begin{figure*}[t]
\centering
\begin{tikzpicture}
  
    \node[inner sep=0pt] (a) at (0,0)
        {\includegraphics[width=0.7\textwidth]{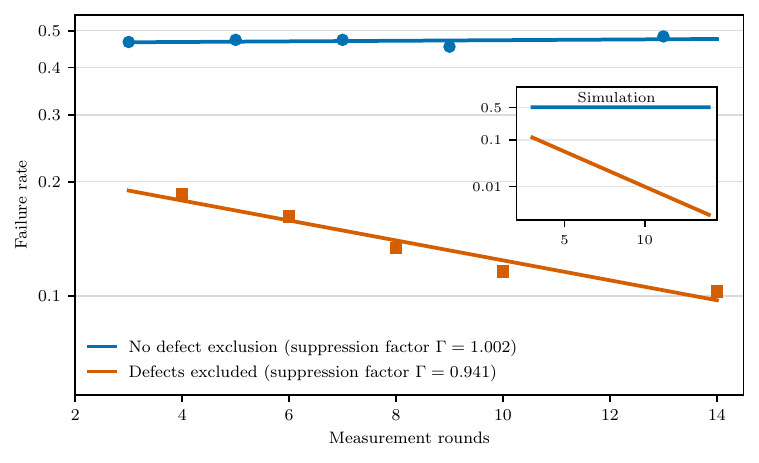}};
    \node[inner sep=0pt, anchor=west, yshift=0.6cm] (b) at ([xshift=0.4cm]a.east)
        {\includegraphics[width=0.3\textwidth]{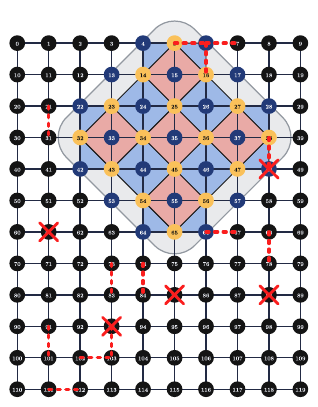}};
    \node[anchor=north west, inner sep=0pt]
        at ([xshift=4pt,yshift=5pt]a.north west) {(a)};
    \node[anchor=north west, inner sep=0pt]
        at ([xshift=2pt,yshift=5pt]b.north west |- a.north west) {(b)\hspace{1em}2026-06-09};

\end{tikzpicture}

   \caption{{\bf Stability experiment without a defective data qubit.} 
Same experiment and analysis as in Fig.~\ref{fig:Miami_stability}, performed on a code without a defective data qubit. The fitted value of $\Gamma$ is comparable to that of the main experiment.}
    \label{fig:Miami_stability2}
\end{figure*}

Figure~\ref{fig:Miami_stability2} presents an additional stability experiment, performed on a code containing some defective couplers and measure qubits, but no defective data qubit, as a comparison with the experiment shown in Figure~\ref{fig:Miami_stability}. From the fit, we obtain a similar value of the suppression factor $\Gamma$, suggesting that the defective data qubit does not significantly affect the stability experiment.

\section{Calibration-based noise model} \label{calib_data}
We build a circuit-level noise model directly from the device calibration data provided by IBM, which publishes per-qubit and per-coupler error rates that are refreshed on a regular schedule. For every experiment we use the calibration snapshot recorded on the same day as the corresponding hardware run.
Each calibrated quantity is mapped onto a Stim noise instruction that is inserted after the associated operation in the syndrome-extraction circuit.
Table~\ref{tab:noise_map} summarises this mapping.

\begin{table}[h!]
\centering
\resizebox{\columnwidth}{!}{%
\begin{tabular}{lll}
\hline
Operation & Calibration quantity & Channel \\
\hline
1-qubit gate  & Pauli-$X$ error $p_{1q}$        & $\mathrm{DEPOLARIZE1}(p_{1q})$ \\
2-qubit gate         & $CZ$ error $p_{CZ}$             & $\mathrm{DEPOLARIZE2}(p_{CZ})$ \\
Measurement          & measure error $p_m$            & $M(p_m)$ \\
Idle                 & $T_1,T_2$, readout length $t$  & $\mathrm{PAULI\_CHANNEL\_1}(p_X,p_Y,p_Z)$ \\
\hline
\end{tabular}%
}
\caption{Mapping from IBM calibration data to Stim noise channels.}
\label{tab:noise_map}
\end{table}

Occasionally, a coupler has no entry in a given day's calibration snapshot. A missing entry is ambiguous: it can indicate a genuinely faulty coupler, or one that is in fact operational but happened to fail characterization. By default, we assign it a high $CZ$ error rate and treat it as defective. When experimental data indicates that the coupler is fully broken, we set its error rate to the maximal two-qubit depolarizing rate ($p_{CZ}=15/16$), whereas if it shows that it is fully functional, we instead assign it the median error rate of the calibrated couplers.

We also incorporate idling errors during the measurement process.  Qubits that are not being actively operated on, most importantly the data qubits while the measure qubits are measured, decohere through amplitude damping ($T_1$) and dephasing ($T_2$). Following Ref.~\cite{Tomita_2014}, we approximate the combined damping process over an idle window of duration $t$ by its Pauli-twirled channel, a single-qubit Pauli channel that applies $X$, $Y$ and $Z$ with probabilities
\begin{equation}
\begin{aligned}
    p_X = p_Y &= \tfrac{1}{4}\bigl(1 - e^{-t/T_1}\bigr), \\
    p_Z       &= \tfrac{1}{2}\bigl(1 - e^{-t/T_2}\bigr)
                 - \tfrac{1}{4}\bigl(1 - e^{-t/T_1}\bigr).
\end{aligned}
\label{eq:idle_pauli}
\end{equation}
The relaxation and coherence times $T_1$ and $T_2$, together with the idle duration $t$ (the reported readout length), are obtained from the calibration data.
For the unmodified code of \cref{fig:memory_experiments_d3}, we found that the decoding performance was improved when using a noise model without idling errors, while for the defect-exclusion circuits, including a uniform idling noise model gave better decoding performance.

\section{No-reset syndrome extraction}
\label{app:no-reset}

Mid-circuit measure qubit resets are not available on the IBM \texttt{ibm\_miami} processor. 
However, recent work has shown that omitting resets need not degrade the performance of logical memory experiments, while for stability experiments it reduces the effective timelike distance but can also remove noise associated with the reset operation~\cite{Geher:2025aa}. 
We employ reset-free syndrome extraction throughout this work.

Without reset, ancilla qubit measurements correspond to cumulative parity rather than the instantaneous stabilizer value. If $m_t$ denotes the measurement outcome in round $t$, then
\begin{equation*}
m_t = s_1 \oplus s_2 \oplus \cdots \oplus s_t,
\end{equation*}
where $s_t$ is the stabilizer value in round $t$. Consequently,
\begin{equation*}
 s_t = m_t \oplus m_{t-1}, 
\end{equation*}
and an error detector comparing consecutive stabilizer values becomes
\begin{equation*}
D_t = s_t \oplus s_{t-1}
     = m_t \oplus m_{t-2}.
\end{equation*}
Bulk detectors therefore compare measurement outcomes separated by two rounds rather than one. Initialization and final-round detectors are modified analogously.

When excluding underperforming components, some stabilizers or super-stabilizers are measured only every other round. As an example, around an excluded measure-qubit in the bulk, the missing weight-four stabilizer is reconstructed as a super-stabilizer: the product of two two-body checks, measured by two neighboring measure qubits $a_1$ and $a_2$. These two checks are read out together, but only once every two rounds, since the intervening round is used to measure the complementary stabilizer type. For such a stabilizer no value is available at round $t-1$, so its detector compares the two nearest rounds in which it is available,
\begin{equation*}
    D_t = s_t \oplus s_{t-2},
\end{equation*}
where $s_t$ is the super-stabilizer value at round $t$. The super-stabilizer value is the product of its two constituent checks, and with no reset each check value is itself the difference of consecutive readouts of its measure qubit, $m^{a_i}_t \oplus m^{a_i}_{t-1}$. Hence,
\begin{equation*}
    s_t = \big(m^{a_1}_t \oplus m^{a_1}_{t-1}\big)
      \oplus \big(m^{a_2}_t \oplus m^{a_2}_{t-1}\big),
\end{equation*}
and the detector becomes an eight-term parity of raw ancilla outcomes,
\begin{equation*}
\begin{aligned}
D_t ={}&
m^{a_1}_t \oplus m^{a_1}_{t-1}
\oplus m^{a_2}_t \oplus m^{a_2}_{t-1}\\
&\oplus m^{a_1}_{t-2}\oplus m^{a_1}_{t-3}
\oplus m^{a_2}_{t-2}\oplus m^{a_2}_{t-3}.
\end{aligned}
\end{equation*}
With reset, each check outcome would directly be the instantaneous check value, $s_t = m^{a_1}_t \oplus m^{a_2}_t$, and the same detector would involve only four terms.

\end{document}